# Interaction Between Superconducting and Ferromagnetic Order Parameters in Graphite-Sulfur Composites

S. Moehlecke[1], Y. Kopelevich[1], and M. B. Maple[2]

[1]Instituto de Física "Gleg Wataghin", Universidade Estadual de Campinas, Unicamp 13083-970, Campinas, São Paulo, Brasil

[2]Dept. of Physics and Institute for Pure and Applied Physics Sciences, UCSD, La Jolla, CA 92093-0360, USA

ABSTRACT

The superconductivity of graphite-sulfur composites is highly anisotropic and associated with the graphite planes. The superconducting state coexists with the ferromagnetism of pure graphite, and a continuous crossover from superconducting to ferromagnetic-like behavior could be achieved by increasing the magnetic field or the temperature. The angular dependence of the magnetic moment m(α) provides evidence for an interaction between the ferromagnetic and the superconducting order parameters.

PACS numbers: 74.10.+v, 74.70.Wz, 74.81.-g

Recently considerable interest in graphite and related materials [1-17] has been triggered by the observation of ferromagnetic (FM) - and superconducting (SC) - like magnetization hysteresis loops M(H) in highly oriented pyrolitic graphite (HOPG) samples even at room temperature [18]. The occurrence of high temperature ferromagnetism has been reported for extraterrestrial graphite [9], proton-irradiated graphite [16], and various carbon-based materials consisting of curved graphite-like sheets [5, 7, 12, 15]. Whereas there is no general consent regarding its origin [1, 2, 6, 13, 14, 17], the so far accumulated experimental evidences indicate that a structural disorder, topological defects, as well as adsorbed foreign atoms can be responsible for the occurrence of both ferromagnetic and superconducting patches in graphitic structures. In particular, it has been demonstrated that sulfur adsorption induces SC [3, 4, 10] in graphite within some sort of "grains" or domains [3, 10]. It has been also demonstrated that the SC in graphite-sulfur (C-S) composites is highly anisotropic and associated with the graphite planes [10]. The SC domains coexist with the FM of pure graphite, and a continuous crossover from SC to FM-like behavior could be achieved by increasing either the applied magnetic field H or the temperature T [10]. An interplay between SC and FM order parameters has been theoretically analyzed for both graphite [1] and C-S composites [11]. However, little is known on this issue from the experimental side.

Here we focus our attention on the highly anisotropic nature of the SC state of the C-S composites, which was explored by means of the angular dependence of the sample magnetic moment $m(\alpha, T, H)$, where $\alpha$ is the angle between the applied magnetic field H and the largest sample surface. The main conclusion of this work is that SC and FM order parameters interact in such a way that the FM component of m(α, T, H) is rotated by $90^o$ below the SC transition temperature $T_c(H)$.

The graphite-sulfur sample studied in this work was thoroughly characterized in Ref. [10]. In summary, the C-S sample was prepared using graphite rods from Carbon of America Ultra Carbon, AGKSP grade, ultra "F" purity (99.9995%) (Alfa-Aesar, # 40766) and sulfur chunks from American Smelting and Refining Co. that are spectrographically pure (99.999+ %). A pressed pellet (ϕ = 6 mm, ~7000 lbf) of graphite was prepared by pressing graphite powder that was produced by cutting and grinding the graphite rod on the edge and side area of a new and clean circular diamond saw blade. The graphite pellet was encapsulated with sulfur chunks (mass ratio ~ 1:1) in a quartz tube under 1/2 atmosphere of argon and heat treated in a tube furnace at 400 °C for one hour and then slowly cooled (4 °C/h) to room temperature. Following this recipe a reasonable reproducibility (~75 %) was reached. X-ray diffraction measurements (θ-2θ geometry and rocking curves) of the reacted sample yielded a spectrum with only the superposition of the (00ℓ) diffraction peaks of graphite with the orthorhombic peaks of sulfur with no extra peak due to a compound, second phase or impurity. The c-axis lattice parameter (c = 6.72 Å) of the sample is equal to the pristine graphite powder pellet, which precludes sulfur intercalation. The diffraction pattern also shows a strong (00ℓ) preferred orientation, which was confirmed by rocking curve scans that yield a Δθ = 6° (FWHM) for the (002) peak, due to the highly anisotropic (plate-like) shape of the graphite grains. The sample (~ 5 x 2.5 x 1.7 $mm^3$) was cut from the reacted pellet and used for the magnetic moment measurements as well as the above described analyses. The magnetic moment m(α, T, H) was measured using a SQUID magnetometer (MPMS5, Quantum Design). All the magnetic moments presented here were normalized to the sample mass. The angular dependence of the magnetic moment of the sample was measured with the MPMS5 magnetometer where a horizontal sample rotator insert (Quantum Design) was placed in the regular sample holder space and controlled by the QD CPU board special EPROM (S3) rotational transport software. The rotator has a substrate capable of rotating 360° around the horizontal axis, and the largest surface of the sample was glued on this substrate with Duco cement in such a way that the graphite c-axis can rotate around the substrate horizontal axis while the applied magnetic field is always vertical and perpendicular to the rotator axis. The measurements were performed with a step of $10^o$. The background signal of the rotator without a sample but with Duco cement is paramagnetic with a susceptibility $\chi_b \sim 3.6 \times 10^{-8}$ emu/Oe. This value is at least one order of magnitude smaller than our sample signal. All the dc magnetic moment measurements were made using a scan length of 1.5 cm.

The SC characteristics of our C-S sample were described in detail in Ref. [10]. Shown in Fig. 1(a) is the temperature dependence of the ZFC (zero field cooled) magnetic moment m(T) after subtraction of the magnetic moment measured at T = 10 K, m(10K); i. e., in the normal state, for several magnetic fields applied perpendicular to the largest surface of the sample (H//c). ZFC measurements were made on heating after the sample was cooled in zero applied field to low temperatures and the desired magnetic field was applied. From Fig. 1(a) it can be seen that the SC transition temperature $T_c$ = 9 K and that the SC signal |m(T) – m(10K)|, increases below this temperature. The m(T) measurements with the applied magnetic field parallel to the main sample surface (H // a) yield a completely different magnetic response. Figure 1(b) shows the temperature dependence of the magnetic moment m(T) measured by ZFC for various applied fields, as indicated in the figure, when H //a. No sign of a SC transition could be detected within the data noise of ~ 5 x $10^{-6}$ emu/g and the range of temperatures measured, down to 2.0 K (not shown in Fig. 1(b)). Note that the scale ranges in Fig. 1(b) and 1(a) are almost the same. These results indicate that the SC state is highly anisotropic and is associated with the graphite planes. Figure 2 shows the hysteresis loops m(H) – $m_o$ (H) measured with the ZFC procedure (H // c) for T = 7, 8, 9 and 10 K, after subtraction of the diamagnetic background signal $m_o = \chi$ H, where $\chi = -7.12 \times 10^{-6}$ emu $g^{-1}$ $Oe^{-1}$ for all these measured temperatures (for more details see [10]). Figure 2(a) shows a characteristic type II SC hysteresis loop at T = 7 K. As the temperature increases above $T_c$ = 9 K, the hysteresis loops resemble those of FM materials (Fig. 2 (c)). For temperatures at and just below $T_c$, the presence of both SC and FM contributions to the measured signal can be seen (Fig. 2 (b,c)).

In other words, the results presented in Fig. 2 provide evidence for the coexistence of SC and FM in the C-S. A similar conclusion has been drawn in Ref. [10] based on the isothermal m(H) measurements at $T < T_c$ throughout a broader field range.

Again, for the m(H) measurements with the applied magnetic field parallel to the graphite planes (H // a), a different magnetic response is obtained. Figure 3(a-c) shows the ZFC magnetic moment hysteresis loops m(H) for T = 5, 7 and 9 K after subtraction of the linear diamagnetic background signal ($m_o = \chi$H, where $\chi$(5 K) = $\chi$(7 K) = $\chi$(9 K) = -2.25 x $10^{-6}$ emu $g^{-1}$ $Oe^{-1}$), (for details, see [10]). From these plots (Figs. 3(a-c)), three almost identical FM-like hysteresis loops, obtained both below and above $T_c$, are clearly seen.

These hysteresis loops are typical of FM materials and are similar to those observed before [18] in HOPG graphite samples. The FM behavior of both HOPG and C-S persists well above room temperature [10, 18]. At the same time, no noticeable change or anomaly was observed in the hysteresis loops around 9 K (H // a). Thus, our samples show a FM-like behavior for all temperatures below the Curie temperature (~750 K) [10] in both H // a and H //c field configurations, and for T < $T_c$(H) and low fields, the FM coexists with a SC state.

We also studied in detail the angular dependence of the magnetic moment m(α) of our sample in both SC and normal states for magnetic fields up to 5000 Oe, using the horizontal sample rotator insert (Quantum Design). The main results are summarized in Figs. 4(a-f) in which are shown the angular dependence of the magnetic moment between 0° ≤ α ≤ 360° for different applied magnetic fields (as indicated in each figure) when the sample is in the SC state (▲, T = 5 K) and the normal state (■, T = 12 K). These measurements were made by initially zero field cooling (ZFC) the sample to 5 K (or 12 K) with α = 0° (H // a).

The angle α given here is the value determined by the QD software and depends only of the initial position. We always tried to align the initial position (α = 0°) of the sample as close as possible so that the magnetic field is parallel to its largest surfaces (H // a). The magnetic moments were measured with increasing and decreasing α from 0° to 360° and back in steps of 10° for H = 30 Oe. Then the magnetic field was raised to H = 50 Oe and m(α) measured again and so on until H = 5000 Oe. Figure 4(a) shows the angular dependences of the magnetic moment m(α) in both the SC state (T = 5 K) and the normal state (T = 12 K) when H = 30 Oe. For T = 5 K the magnetic moments measured with increasing (▲) and decreasing (▼) α are also shown, demonstrating the reversibility of m(α). In the other figures only the measurements with increasing α are presented. The applied magnetic field direction is shown in Fig. 4(a). At the top of Fig. 4(a), we also show schematically the sample cross section orientation with respect to the magnetic field for $\alpha$ values of ~ 0° (H // a), 90° (H // c), 180° (H // a), 270° (H // c) and 360° (H // a); these same sample/field configurations are valid for the other figures (Fig. 4(b-f)).

Several new and interesting observations can be made from these measurements. From Fig. 4(a), it can be seen that m($\alpha$) ~ sin $\alpha$ for T = 12 K and m($\alpha$) ~ sin ($\alpha$ + 90°) = cos $\alpha$ for T = 5 K. These oscillatory behaviors imply that the magnetic response of the sample can vary from paramagnetic (or ferromagnetic) to diamagnetic depending simply on the sample/field configuration used during the measurement. The m($\alpha$, T, H) behavior of this graphite-sulfur sample in the normal state, T = 12 K, is essentially the same as found in pure HOPG graphite samples, which will be reported elsewhere [19]. The m($\alpha$) dependence for T = 5 K of Fig. 4(a) suggests that the paramagnetic (or ferromagnetic) m(H, T, $\alpha$) is confined to the graphite planes (or parallel to the largest surfaces of the sample) and rotates with the sample in the applied field. For T = 12 K (i.e., in the normal state), the magnetic moment is also paramagnetic (or ferromagnetic) but is perpendicular to the graphite planes (90° out of phase as compared to the measurements performed at T = 5 K) and also rotates with the sample. We have shown before that this sample exhibits superconducting behavior for T = 5 K ($T < T_c$). It should be noted, however, that at H = 30 Oe and T = 5 K the SC diamagnetic signal is only ~ 10 % (see Fig. 1(a)) of the magnetic moment values of Fig. 4(a). Thus, at 5 K the SC signal is masked by a larger FM moment. Also, the “diamagnetic” signals observed in the normal state for H // c and H // a [10] as shown in Fig. 1(b) (H // a), are, in fact, a result of the superposition of the FM moment pointing against the applied magnetic field and/or when 180° ≤ $\alpha$ ≤ 340° in Fig. 4(a) (T = 12 K). As the magnetic field is increased, at both T = 5 K and 12 K, m($\alpha$) begins to change behavior for H of the order of 100-300 Oe (see Figs. 4(c) and (d)). For high fields like H = 1000 Oe (Fig. 4(e)) and higher (H = 5000 Oe, Fig. 4(f)) the two magnetic moment dependences tend to almost identical behavior and scale as $m(\alpha) \propto \sin 2\alpha$ (a function that repeats itself every 180°). These are due to the Landau diamagnetism associated with the graphite planes [20] that begin to dominate at these high magnetic fields. The results of Figs. 4(a-f) show that at low fields, $m(\alpha) \propto \sin \alpha$ (or $\sin \alpha + 90°$) and, as the magnetic field increases, it changes to $m(\alpha) \propto \sin 2\alpha$. Thus, a function of the form;

$$m(\alpha) = m_o + A_\alpha \sin(\alpha + \phi_\alpha) + A_{2\alpha} \sin(2\alpha + \phi_{2\alpha}) \qquad (1)$$

should describe all types of m(α, H, T) behavior. The continuous lines through the experimental data of figures 4(a-f) are fits with this function (1), which also describes well the m(α, H, T) dependences of pure HOPG samples [19].

It is reasonable to assume that the main cause of the different m(α) behaviors between T = 5 K and 12 K at low fields originates from the occurrence of SC below 9 K. In spite of the fact that total SC signal is rather weak [10], the experimental results presented in this paper provide evidence for an interaction between SC and FM order parameters turning the FM moment direction by 90° and confining it in the graphite planes. One may speculate on the appearance at $T < T_c$ of spin-polarized currents associated with spin-triplet SC [1, 11]. In this case, these currents would exert a torque on the preexisting FM moment, leading to its rotation [21, 22]. Certainly, future work is needed to examine this conjecture.

This work was partially supported by FAPESP, CNPq and the US Department of Energy under Grant DE-FG03-86ER-95230.

Figure Captions

Figure 1. (a) Temperature dependences of the SC diamagnetic moment measured by zero-field-cooling after subtraction of the normal magnetic moment at 10 K, m(10K), at various fields: (▲), H = 10 Oe; (★), H = 30 Oe; (○), H = 50 Oe; (◊), H = 70 Oe; (○), H = 100 Oe; (∇), H = 150 Oe; (○), H = 300 Oe, with H // c. (b) Temperature dependences of the magnetic moment measured by zero-field-cooling with H // a and at different magnetic fields, as indicated in oersteds next to each curve.

Figure 2. Zero-field-cooled magnetic moment hysteresis loops m(H), after the subtraction of the diamagnetic background signal ($m_o = \chi H$, where $\chi = -7.12 \times 10^{-6}$ emu $g^{-1}$ $Oe^{-1}$), with H // c and for (a) T = 7 K, (b) T = 8 K, (c) T = 9 K and (d) T = 10 K. For details see text.

Figure 3. Zero-field-cooled hysteresis loops m(H), after the subtraction of the diamagnetic background signal ($m_o = \chi H$, where $\chi = -2.25 \times 10^{-6}$ emu $g^{-1}$ $Oe^{-1}$), with H // a and for (a) T = 5 K, (b) T = 7 K and (c) T = 9 K. For details see text.

Figure 4. Angular dependences of the magnetic moment $m(\alpha)$ between 0° and 360° in the SC state (T = 5 K, (▲)) and in the normal state (T = 12 K, (■)) at various magnetic fields: (a) H = 30 Oe; (b) H = 50 Oe; (c) H = 100 Oe; (d) H = 300 Oe; (e) H = 1000 Oe; (f) H = 5000 Oe. The applied magnetic field direction is shown in all figures 4(a-f). At the top of figure 4(a) is also shown schematically the sample cross section position in relation to the magnetic field for $\alpha$ values of ~ 0° (H // a), 90° (H // c), 180° (H // a), 270° (H // c) and 360° (H // a), these same sample/field configurations are valid for the other figures 4(b-f).

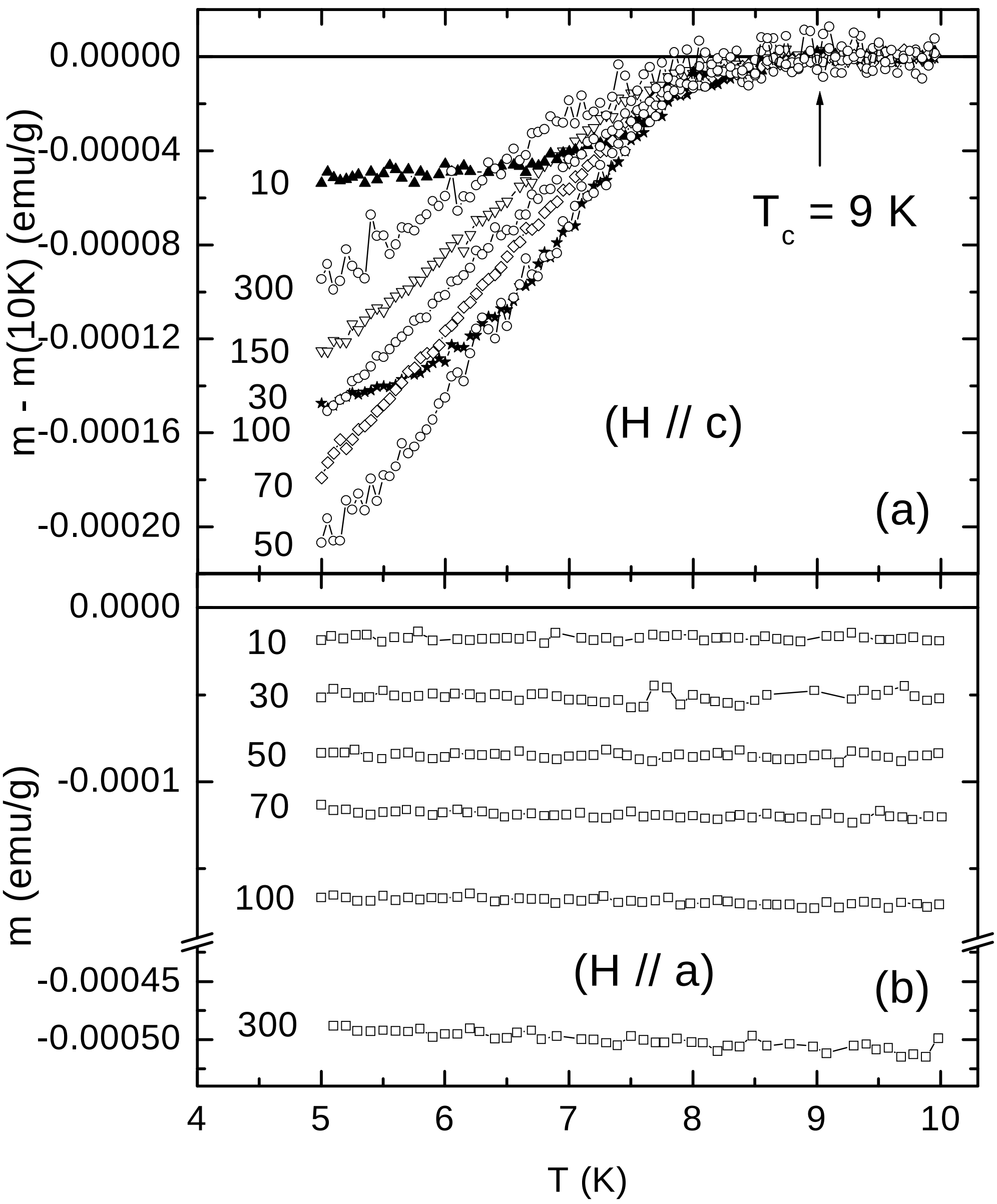

Fig. 1

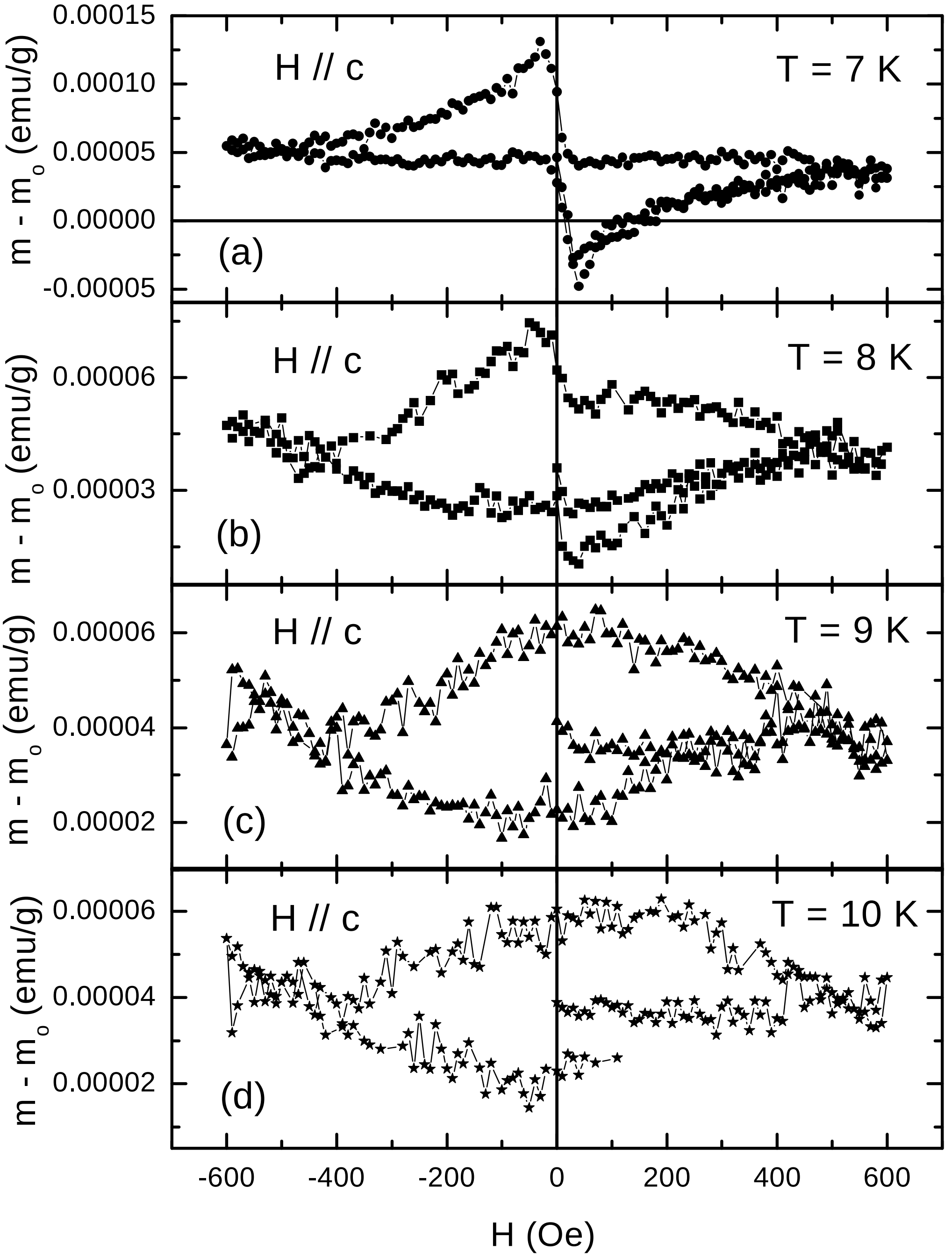

Fig. 2

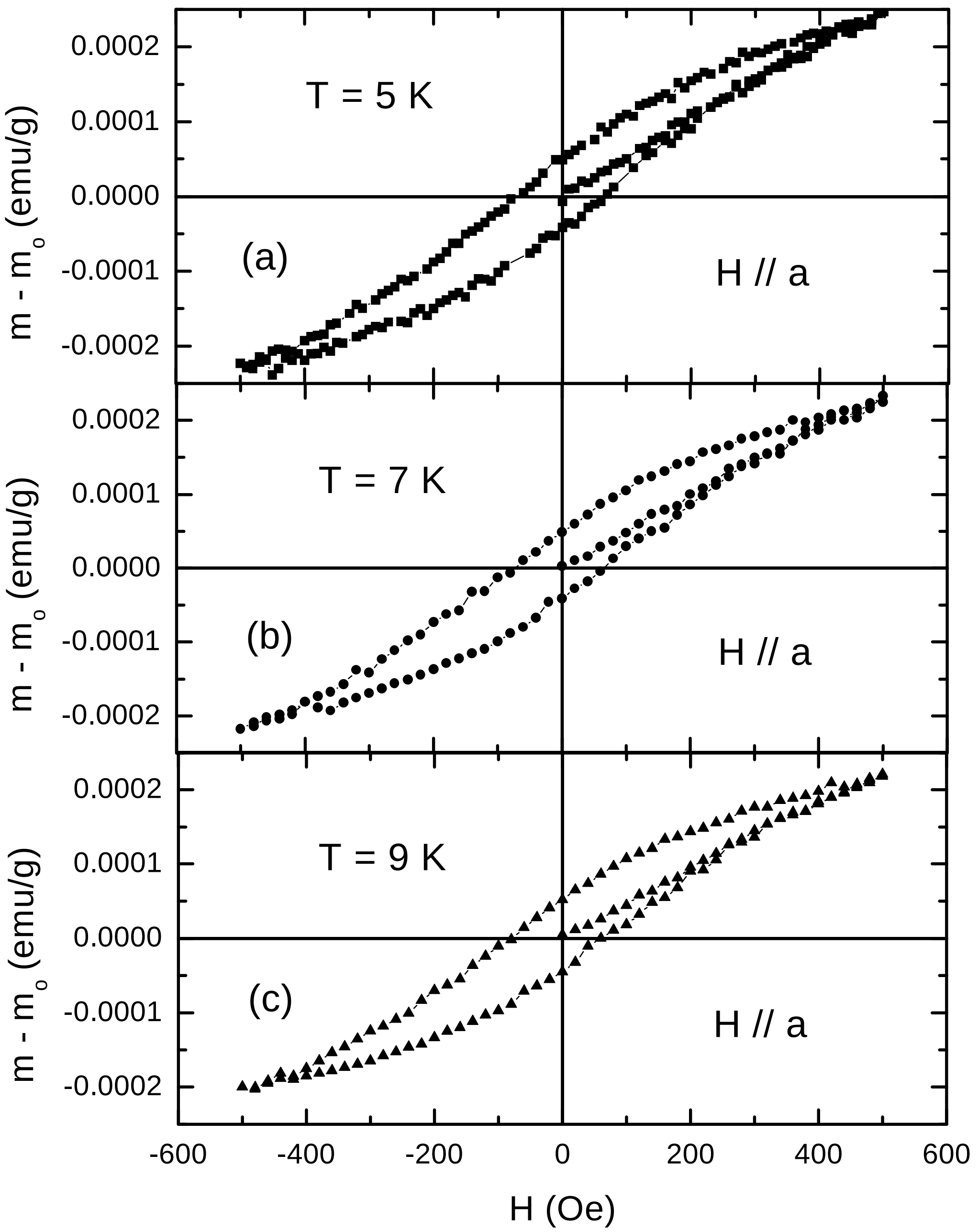

Fig. 3

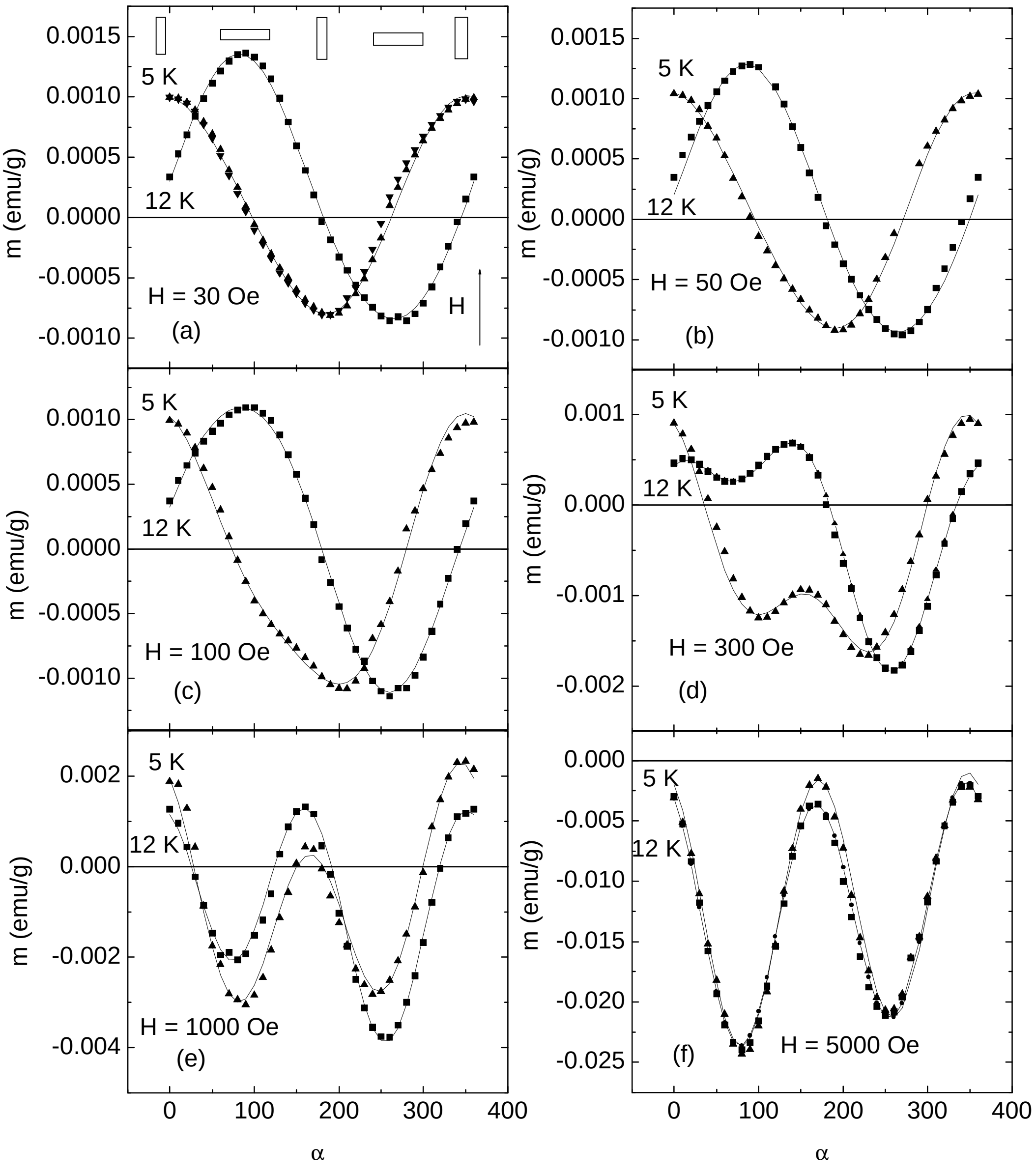

Fig. 4